\documentclass{article}
\newcommand{\bb}{\begin{eqnarray}}
\newcommand{\ee}{\end{eqnarray}}
\begin{document}
\title{ Non-commutative oscillators and the commutative limit}
\author{B. Muthukumar\thanks{e-mail muthu@theory.saha.ernet.in}\\ 
and\\
P. Mitra\thanks{e-mail mitra@theory.saha.ernet.in}\\
Saha Institute of Nuclear Physics\\
Block AF, Bidhannagar\\
Calcutta 700 064, INDIA}
\date{hep-th/0204149}
\maketitle
\begin{abstract}
It is shown in first order perturbation theory
that anharmonic oscillators in non-commutative
space behave smoothly in the commutative limit
just as harmonic oscillators do.
The non-commutativity provides a method for
converting a problem in degenerate perturbation
theory to a non-degenerate problem.
\end{abstract}

In the last few years theories in non-commutative space \cite{1}-\cite{5}
have been studied extensively. While the motivation for this kind of space
with non-commuting coordinates is mainly theoretical, it is possible
to look experimentally for departures from the usually assumed 
commutativity among the space coordinates \cite{9}-\cite{11}. So far no clear 
departure has been found, but it is clear that any experiment can only provide
a bound on the amount of non-commutativity, and that more precise
experiments in future can reveal a small amount. Meanwhile,
there are some theoretical issues which have arisen in the course
of these investigations.
If one calls the non-commutativity parameter $\theta$, so that two
spatial coordinates $\mathbf{x},\mathbf{y}$ satisfy the relation
\[
[\mathbf{x},\mathbf{y}]=i\theta,
\]
one would expect ordinary commutative space to emerge in the limit
$\theta\to 0$. In many field
theoretical and quantum mechanical problems, however, the passage from the
non-commutative space to its commutative limit has {\it not} appeared to be
smooth \cite{12}-\cite{14}. The literature is replete with expressions 
where $\theta$ appears in the denominator. The simplest system is 
the two-dimensional harmonic oscillator. As in commutative
space, this quantum mechanical problem is again exactly solvable 
\cite{annals}-\cite{ss}, and
the spectrum is in fact smooth in the commutative limit, 
but the literature is not very clear about the situation: there seems to be 
a lack of smoothness in the generic case \cite{15,16}. 
For a clarification of this limit, 
we will first review the two-dimensional harmonic oscillator 
and then go over to a perturbation 
$(\mathbf{x}^{2}+\mathbf{y}^{2})^{2}$ to see if the smoothness survives.
The anharmonic problem cannot be solved exactly
even in commutative space. But a perturbative treatment shows that the
quartic terms do
not destroy the smoothness of the $\theta\to 0$ limit.
It is interesting to note that the commutative oscillator here involves 
a degenerate perturbation problem, while the non-commutative one is 
non-degenerate.

Let us write the two dimensional anharmonic oscillator potential in the form
\begin{eqnarray}
\frac{1}{2} m \omega^{2} (\mathbf{x}^{2}+\mathbf{y}^{2}) 
+\alpha (\mathbf{x}^{2}+\mathbf{y}^{2})^{2}.
\end{eqnarray}
The non-commuting coordinates can be expressed in terms of commuting 
coordinates and their momenta in the form
\begin{eqnarray}
\mathbf{x}& = & x-\frac{1}{2\hbar}\theta p_y, \nonumber \\
\mathbf{y}& = & y+ \frac{1}{2\hbar}\theta p_x. 
\end{eqnarray}
The Hamiltonian for the unperturbed system is 
\begin{eqnarray}
H_{HO} & = & \frac{1}{2m} \left( p^{2}_{x} + p^{2}_{y}
\right) + \frac{1}{2} m \omega^{2} \left( \mathbf{x}^{2} + 
\mathbf{y}^{2}\right)\nonumber\\
& =& \frac{1}{2m} \left( p^{2}_{x} + p^{2}_{y} \right)
+ \frac{1}{2}m\omega^{2} \left( \left( x-\frac{1}{2\hbar} \theta
p_y\right)^{2} + \left( y + \frac{1}{2\hbar} \theta p_{x}^{2}
\right)^{2}\right)\nonumber\\
& = & \left( \frac{1}{2m}+\frac{m\theta^{2}\omega^{2}}{8
\hbar^{2}}\right)\left( p_x^{2}+p_y^{2}\right)+
\frac{1}{2}m\omega^{2}\left( x^{2}+y^{2}\right) \nonumber\\
& & -\frac{m\omega^{2}\theta}{2\hbar}\left( x p_y-yp_x\right). 
\end{eqnarray}
It is convenient to set $\left( \frac{1}{2m} +
\frac{m\theta^{2}\omega^{2}}{8\hbar^{2}}\right)\equiv\frac{1}{2M}$ and 
$m\omega^{2}\equiv M\Omega^{2}$.
One can introduce the ladder operators through the equations
\begin{eqnarray}
a_x = \sqrt{\frac{M\Omega}{2\hbar}}\left(x+\frac{ip_x}{M\Omega}\right)
&;&\;\;a_x^{\dagger} =
 \sqrt{\frac{M\Omega}{2\hbar}}\left(x-\frac{ip_x}{M\Omega}\right);\nonumber\\
a_y =
 \sqrt{\frac{M\Omega}{2\hbar}}\left(y+\frac{ip_y}{M\Omega}\right)
&;&\;\;a_y^{\dagger}
=\sqrt{\frac{M\Omega}{2\hbar}}\left(y-\frac{ip_y}{M\Omega}\right); \nonumber\\
x=\sqrt{\frac{\hbar}{2M\Omega}}\left(a_x+a_x^{\dagger}\right)&;&\;\;
 p_x= \frac{1}{i}\sqrt{\frac{M\Omega
 \hbar}{2}}\left(a_x-a_x^{\dagger}\right);\nonumber\\
y=\sqrt{\frac{\hbar}{2M\Omega}}\left(a_y+a_y^{\dagger}\right)&;&\;\;
 p_y= \frac{1}{i}\sqrt{\frac{M\Omega
 \hbar}{2}}\left(a_y-a_y^{\dagger}\right).
\end{eqnarray}
In terms of these operators the unperturbed Hamiltonian takes the form
\begin{equation}
H_{HO}=\hbar \Omega \left( a_x^{\dagger}a_x+a_y^{\dagger}a_y+1\right)-
\frac{M\Omega^{2}\theta}{2i}\left(a_x^{\dagger}a_y
-a_y^{\dagger}a_x\right).
\end{equation}
In view of the Schwinger representation for the angular momentum, 
\begin{eqnarray}
J_1 &=&\frac{1}{2}\left(a_x^{\dagger}a_y + a_y^{\dagger}a_x\right),\nonumber\\
J_2 &=&\frac{1}{2i}\left(a_x^{\dagger}a_y - a_y^{\dagger}a_x\right),\\
J_3 &=&\frac{1}{2}\left(a_x^{\dagger}a_x - a_y^{\dagger}a_y\right),\nonumber
\end{eqnarray}
the part $a_x^{\dagger}a_y - a_y^{\dagger}a_x$ in $H_{HO}$ is seen to be
$2iJ_2$.
Under the unitary transformation
\begin{equation}
 \left(\begin{array}{c}
        a_x \\ a_y 
       \end{array} \right) = \frac{1}{\sqrt{2}}
                              \left(\begin{array}{cc}
                                 1 & -i \\ i & -1 
                              \end{array} \right)
                              \left(\begin{array}{c}
                                 a_x' \\ a_y'
                              \end{array} \right),   \label{unitary}
\end{equation}                      
in which this piece takes the diagonal form $2iJ_3'$, the
Hamiltonian becomes
\begin{equation}
H_{HO}=\hbar \Omega \left( a_x'^{\dagger}a_x' + a_y'^{\dagger}a_y'
+1\right)-\frac{M\Omega^{2}\theta}{2}\left(a_x'^{\dagger}a_x'
-a_y'^{\dagger}a_y' \right).
\end{equation}
If $\hat{N}_x=a_x'^{\dagger}a_x$ and $\hat{N}_y=a_y'^{\dagger}a_y$ are
the number operators of the harmonic oscillators in the $x$ and $y$
directions respectively, one can write
\begin{equation}
H_{HO}=
\hbar\Omega\left(\hat{N}_x+\hat{N}_y+1\right)
-\frac{M\Omega^{2}\theta}{2}\left(\hat{N}_x-\hat{N}_y\right).
\end{equation}
The eigenvalues of the unperturbed Hamiltonian are therefore
\begin{equation}
E^0_{n_x,n_y}= \hbar \Omega
\left(n_x+n_y+1\right)-\frac{M\Omega^{2}\theta}{2}\left(n_x-n_y\right),
\end{equation}
where $n_x,n_y$ are non-negative integers.
In terms of $m$ and $\omega$, the eigenvalues can be written as
\begin{equation}
E^0_{n_x,n_y}=m\omega^{2}\hbar^{2}\left(\frac{1}{m^2 \omega^2
\hbar^2}+\frac{\theta^2}{4\hbar^4}\right)^{\frac{1}{2}}\left(n_x+n_y+1\right)
-\frac{m\omega^2\theta}{2}\left(n_x-n_y\right).
\end{equation}
These eigenvalues are generically non-degenerate.
In the limit $\theta\rightarrow 0$ they smoothly 
reduce to the standard degenerate expression
\begin{equation}
E^0_{n_x,n_y}\rightarrow\hbar\omega\left(n_x+n_y+1\right).
\end{equation}
Let us now introduce a perturbation of the form $\alpha 
\left(\mathbf{x}^{2} +\mathbf{y}^{2}\right)^2$.
In terms of the ladder operators,
\begin{eqnarray}
\mathbf{x}&=&\sqrt{\frac{\hbar}{2M\Omega}}\left(a_x+a_x^{\dagger}\right)
-\frac{1}{i}\sqrt{\frac{M\Omega\theta^2}{8\hbar}}\left(a_y
-a_y^{\dagger}\right),\nonumber\\
\mathbf{y}&=&\sqrt{\frac{\hbar}{2M\Omega}}\left(a_y+a_y^{\dagger}\right)
+\frac{1}{i}\sqrt{\frac{M\Omega\theta^2}{8\hbar}}\left(a_x-a_x^\dagger\right).
\end{eqnarray}
But under the unitary transformation (\ref{unitary}),
\begin{eqnarray}
\mathbf{x} &=&
\frac{1}{2}\left(\sqrt{\frac{\hbar}{M\Omega}}-\sqrt{\frac{M\Omega
\theta^{2}}{4\hbar}}\right)a_x'^{\dagger}+\frac{i}{2}\left(\sqrt{\frac{
\hbar}{M\Omega}}+\sqrt{\frac{M\Omega\theta^{2}}{4\hbar}}\right)a_y'^{\dagger},
\nonumber\\ 
&=& \beta\left(a_x'+a_x'^{\dagger}\right)-i\gamma\left(a_y'
-a_y'^{\dagger}\right),
\end{eqnarray}
where
$\beta=\frac{1}{2}\left(\sqrt{\frac{\hbar}{M\Omega}}
-\sqrt{\frac{M\Omega\theta^2}{4\hbar}}\right)$ and $\gamma
=\frac{1}{2}\left(\sqrt{\frac{\hbar}{M\Omega}}
+\sqrt{\frac{M\Omega\theta^2}{4\hbar}}\right)$.
One also has
\begin{eqnarray}
\mathbf{y}&=& i \beta
\left(a_x'-a_x'^{\dagger}\right)-\gamma\left(a_y'+a_y'^{\dagger}\right).
\end{eqnarray}
Hence,
\begin{eqnarray}
\left(\mathbf{x}^2+\mathbf{y}^2\right)&=&4\beta^{2}a_x'^{\dagger}a_x'
+4\gamma^{2}
a_y'^{\dagger}a_y'+
2 \left(\beta^{2}+\gamma^{2}\right)\nonumber\\
&& \;\;\;-4i\beta\gamma a_x'a_y'+4i\beta\gamma a_x'^{\dagger}a_y'^{\dagger}.
\end{eqnarray}
This is a non-diagonal operator in the basis in which the unperturbed
eigenstates of the Hamiltonian are diagonal, but its effect on the
eigenvalues can be studied in first order perturbation theory.
In view of the non-degeneracy of the unperturbed eigenvalues, it is
sufficient to calculate
the expectation values of $\left(\mathbf{x}^2+\mathbf{y}^2\right)^2$ in the 
states
$\left|n_x,n_y\right>$. Thus,
\begin{eqnarray}
\alpha\left<n_x,n_y\right|\left(\mathbf{x}^2+\mathbf{y}^2
\right)^2\left|n_x,n_y\right>&=& \alpha\left(16
\beta^4n_x(n_x+1)+16\gamma^4n_y(n_y+1)\nonumber\right.\\
&&+32\beta^2\gamma^2(n_x+n_y)+64\beta^2\gamma^2n_xn_y\nonumber\\
&&+4\left.(\beta^4+\gamma^4)+24\beta^2\gamma^2\right).
\end{eqnarray}
The expression on the right hand side
gives the first order correction to the eigenvalues caused by the 
anharmonicity. In the $\theta\to 0$ limit, $\beta,\gamma$ behave smoothly
and this correction goes over smoothly to a finite value:
\begin{equation}
E^1_{n_x,n_y}\rightarrow
\alpha({\hbar\over m\omega})^2(n_x^2+n_y^2+4n_xn_y+3n_x+3n_y+2).\label{corr}
\end{equation}
The smooth transition should make it obvious that the same correction
must be obtained in the case of the commutative anharmonic oscillator.
However, the eigenvalues of the unperturbed commutative oscillator
are degenerate, so it may be more convincing if the agreement is shown
explicitly after doing a degenerate perturbation theory calculation.

In the commutative case $\theta=0$, both 
$\beta,\gamma$ reduce to ${1\over 2}\sqrt{\hbar\over m\omega}\equiv\beta_0$.
The operator of interest is
\begin{equation}
\left({x}^2+{y}^2\right)=4\beta_0^{2}(a_x'^{\dagger}a_x'
+ a_y'^{\dagger}a_y'+ 1 -ia_x'a_y'+i a_x'^{\dagger}a_y'^{\dagger}),
\end{equation}
where the unitarily transformed ladder operators are used for ease
of comparison with the previous calculation.
The matrix element of the square of this operator has to be calculated
between degenerate eigenstates of the unperturbed Hamiltonian. 
Any choice of basis for the degenerate states is permissible: it
is convenient to use the eigenstates of $a_x'^{\dagger}a_x',a_y'^{\dagger}a_y'$
instead of the untransformed number operators. Thus,
states with different values of $n_x,n_y$ but {\it the same value of} 
$n_x+n_y$ are to be considered.
Now $({x}^2+{y}^2)^2|n_x,n_y>$ contains the states
$|n_x,n_y>,|n_x-2,n_y-2>,|n_x+2,n_y+2>,|n_x-1,n_y-1>,|n_x+1,n_y+1>$.
Out of these, only $|n_x,n_y>$ has the original value of $n_x+n_y$,
while all the other states have different values. Thus, although
the operator of interest is not completely diagonal in the energy basis, it
is diagonal in the subspace of states with fixed $n_x+n_y$.
The diagonal value is
\begin{eqnarray}
16\beta_0^4[(n_x+n_y+1)^2+(n_x+1)(n_y+1)+n_xn_y]&&\nonumber\\=
\left({\hbar\over m\omega}\right)^2(n_x^2+n_y^2+4n_xn_y+3n_x+3n_y+2).&&
\end{eqnarray}
These diagonal values are also the eigenvalues of the matrix,
so that the correction to the degenerate unperturbed energy eigenvalue
is $\alpha\left({\hbar\over m\omega}\right)^2(n_x^2+n_y^2+4n_xn_y+3n_x+3n_y
+2)$, which agrees with the $\theta\to 0$ limit (\ref{corr})
of the non-commutative
calculation. One could also carry out the calculation in the untransformed
occupation number basis, where the matrix in the space
of the degenerate eigenvectors is not diagonal to begin
with, but on diagonalization, the same eigenvalues are obtained.
 
Thus, not only the exactly solvable harmonic oscillator but even 
the first order perturbation theory result for the eigenvalues
of the two-dimensional non-commutative anharmonic oscillator
behave smoothly in the commutative limit. It is conceivable
that, as is widely believed, the Coulomb problem may not show this
smoothness. But there clearly is a class of Hamiltonians, not just
an isolated Hamiltonian, whose eigenvalues have a smooth $\theta$-dependence.

A by-product of this demonstration is the emergence 
of a method of handling the degenerate perturbation theory
through conversion to a non-degenerate problem. This happens through the
introduction of the parameter $\theta$, which can be regarded as
a mathematical trick from the point of view of commutative theory.
The calculation may be done for non-zero $\theta$ and then the
limit $\theta\to 0$ taken.

\medskip

We would like to thank Ashok Chatterjee for his questions about
the $\theta\to 0$ limit.

\end{document}